\def\year{2016}\relax
\documentclass[letterpaper]{article}
\usepackage{aaai16}
\usepackage{times}
\usepackage{helvet}
\usepackage{courier}
\usepackage{graphicx}

\usepackage[margin=1in]{geometry} 
\usepackage{amsmath,amsthm,amssymb}

\begin{document}

\title{Dynamic Allocation of Crowd Contributions for Sentiment Analysis \\during the 2016
  U.S. Presidential Election}

\author{
Mehrnoosh Sameki, Mattia Gentil, Kate K. Mays, Lei Guo, and Margrit Betke$^*$\\
\\
{\bf \large Boston University}\\
}

\maketitle

\begin{abstract}
  Opinions about the 2016 U.S. Presidential Candidates have been expressed in
  millions of tweets that are challenging to analyze automatically.
  Crowdsourcing the analysis of political tweets effectively is also difficult,
  due to large inter-rater disagreements when sarcasm is involved.  Each tweet
  is typically analyzed by a {\em fixed} number of workers and majority voting.
  We here propose a crowdsourcing framework that instead uses a {\em dynamic}
  allocation of the number of workers.  We explore two dynamic-allocation
  methods: (1) The number of workers queried to label a tweet is computed {\em
    offline} based on the predicted difficulty of discerning the sentiment of a
  particular tweet.  (2) The number of crowd workers is determined {\em online,}
  during an iterative crowd sourcing process, based on inter-rater agreements
  between labels. We applied our approach to 1,000 twitter messages about the
  four U.S. presidential candidates Clinton, Cruz, Sanders, and Trump, collected
  during February 2016.  We implemented the two proposed methods using decision
  trees that allocate more crowd efforts to tweets predicted to be sarcastic.
  We show that our framework outperforms the traditional static allocation
  scheme.  It collects opinion labels from the crowd at a much lower cost while
  maintaining labeling accuracy.
\end{abstract}

\section{Introduction}

During the 2016 U.S. presidential primary election season, the political debate
on Twitter about the four presidential candidates Hillary Clinton, Ted Cruz,
Bernie Sanders, and Donald Trump was particularly lively and created a huge
corpus of data.  It has been argued that Twitter can be considered a valid
indicator of political opinion~\cite{TumasjanSpSaWe10}, and so various parties,
including journalists, campaign managers, politicians, and social scientists,
are interested in using automated natural language processing tools to mine this
corpus.  

Unsupervised learning methods have been used previously to analyze a similar
corpus, 77 millions tweets about the 2012 U.S. presidential election and create
summary statistics such as ``twitter users mentioned foreign affairs in
connection with Obama more than with Romney'' \cite{GuoVaPaDiIs16}.  Supervised
learning methods also have been used, for example, to analyze filtered
``snippets'' of political blogs~\cite{HsuehMeSi09}.  However, creating {\em
  accurate} learning methods to analyze positive or negative sentiments is
challenging. Political opinions expressed on the internet often contain sarcasm
and mockery~\cite{GuoVaPaDiIs16,HsuehMeSi09}, which are difficult to discern by
machine or human computation~\cite{GonzalezMuWa11,YoungSo12}

Crowdsourcing has been proposed to collect training data for predictive models
used to classify political sentiments~\cite{HsuehMeSi09,WangCaKaBaNa12}.  Out of
concern for the accuracy of human annotation, it is standard practice to collect
multiple labels for the same data point and then use the label that obtained a
majority vote~\cite{KargerOhSh13}.  Typically an odd number of crowd workers,
e.g., five or seven, is chosen to create this redundancy.  Redundancy, however,
cannot guarantee reliability, i.e., agreement among the raters with each other
about the sentiment present in the text in question.  For example, when five
crowd workers analyzed the sentiments expressed in the political snippets
dataset~\cite{HsuehMeSi09}, only a 47\% agreement rate on the three labels
``positive,'' ``negative,'' or ``neutral sentiment'' could be achieved.

Hsueh et al.,~\citeyear{HsuehMeSi09}, noted that ``not all snippets [of
political blogs] are equally easy to annotate.''  We made the same observation
for our data -- sarcastic twitter messages are more difficult to label, and we
propose to allocate crowd resources according to the predicted difficulty level:
The more difficult the sentiment analysis may be, the higher the number of
workers becomes that our model assigns.  In allocating fewer crowd workers to
tasks that are predicted to be easy, we aim to balance the goals of labeling
accuracy and efficiency.

The literature describes techniques for optimal trade-offs between accuracy and
redundancy in crowdsourcing~\cite{KargerOhSh13,TranVeRoJe13}.  In these
works, the proposed crowdsourcing mechanism uses a fixed number of crowd workers
per task, and the assignment is agnostic about the latent difficulty level of
each task.  If the difficulty of a task can be discerned, easy tasks could be
routed to novice workers and difficult tasks to expert
annotators~\cite{KolobovMaWe13}.  Optimal task routing, however, is an NP-hard
problem, and so online schemes for task-to-worker assignments have been proposed
~\cite{BraggKoMaWe14,RajpalGoMa15}.  Our work falls into this category of online
crowdsourcing methodology.

\noindent
Our contributions are as follows:
\begin{itemize}
\item We propose a decision-tree approach for {\em dynamically} determining the
  number of crowd workers for tasks that require redundant annotations.

\item We provide two versions of this approach: The {\em offline} version
  computes the number of workers needed based on the content of the data they
  are asked to analyze.  The {\em online} version relies on iterative rounds of
  crowdsourcing and determines the number based on content and annotation
  results in previous rounds.

\item To illustrate and evaluate our approach, we conducted a crowdsourcing
  experiment with a dataset of 1,000 tweets that were sent during the 2016
  primary election season.  We collected 5,075 ratings of the sentiment towards
  presidential candidates Clinton, Cruz, Sanders, and Trump in these tweets and
  evaluated their accuracy with respect to a gold standard established by
  experts in political communication.

\item Comparisons with traditional crowdsourcing strategies show that the
  proposed offline and online selection methods intelligently detect ambiguities
  in sentiment analysis and recruit more workers to resolve those.  We show that
  a large portion of the crowdsourcing budget can be saved at a small loss of
  accuracy.
\end{itemize}

\section{Method}

We here describe our method to solve the problem of dynamically assigning
crowd workers to analyze the sentiment of political tweets.  Our approach
consists of three main components.  First, we designed a method to detect
sarcasm in tweets (Section~\ref{sec:sarcasm}).  This first step was important
because sarcasm is one of the most confusing and misleading language features to
classify even for a human annotator, especially when a single out-of-context
tweet is being analyzed.

We then constructed a decision tree that assigns to each tweet a fixed number of
crowd workers based on the presidential candidates mentioned in the tweet and
other text properties, in particular, its sarcasm
(Section~\ref{sec:decision-tree}).  In designing such a tree, we were motivated
by the following insight: For tweets which are expected to be clear and
straight-forward to analyze, fewer annotators would be required than for tweets
that are sarcastic and complicated.  To build the tree, we estimated how
troublesome it would be for a crowd worker to correctly understand what kind of
sentiment is being expressed towards the candidates.

The third component of our approach moves from an offline to an online
consideration of how many crowd workers to involve in the labeling process
(Section~\ref{sec:dynamic-allocation}).  Based on the inter-rater agreements
between labels obtained in a first phase of an iterative crowd sourcing process,
for tweets which proved to be challenging to annotate, our method determines how
many additional labels to acquire in one or more subsequent crowd sourcing
phases.

Our final methodological contribution is a description of the equivalency
between two crowdsourcing schemes, the traditional 5-worker-per-task scheme and
the dynamic scheme that assigns 3 workers per task in the first round and 2
additional workers in a second round if disagreement is encountered in the 1st
round.  This is a general result about offline versus online crowdsourcing
schemes.  It holds for any application and is therefore presented in
Section~\ref{sec:off-vs-on}, separate from the results of our sentiment analysis
of political tweets.

\begin{figure*}[t]
  \centering
  \includegraphics[width=1\linewidth]{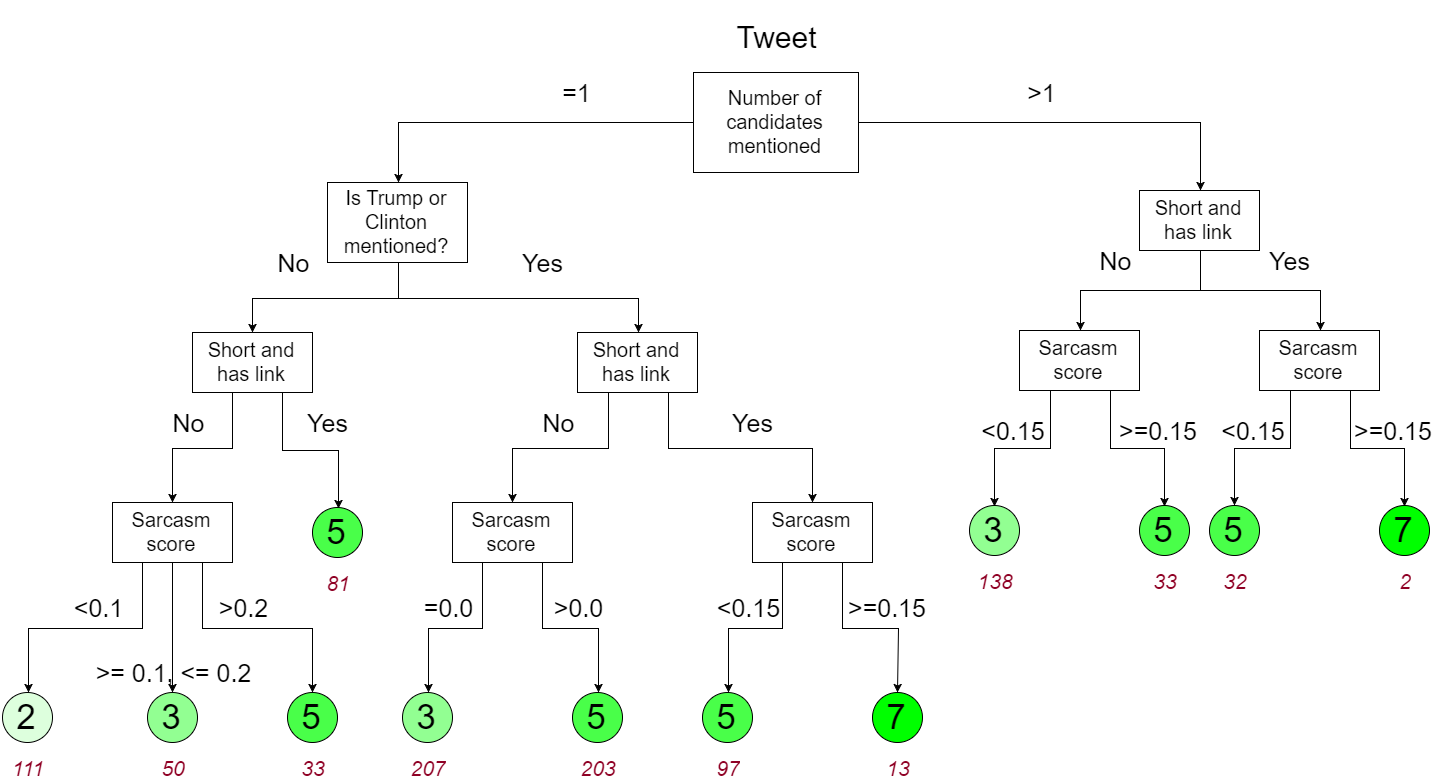}
  \caption{The Static Decision Tree (SDT) model used to determine the number of
    crowd workers (leaves) to engage in analyzing tweets about four presidential
    election candidates.  The intensity of the leaf shading visualizes costs,
    e.g. pale green corresponds to low costs. The sarcasm score is computed
    according to Eq.~\ref{eq:sarcasm-score}.  Experimental results are shown
    under each leaf as the number of tweets processed (red). 
%MEHRNOOSH: MAYBE SMALLER AND CHANGE OF COLOR CODE??
}
\label{figure:decision-tree-diagram}
\end{figure*}

\subsection{Sarcasm Detection}
\label{sec:sarcasm}

Our first step was trying to predict whether a given tweet was sarcastic or not.
We used a Bayesian approach to estimate the likelihood of sarcasm based on
training data provided by domain experts.  Our training data contains the label
``sarcasm present'' or ``sarcasm not present'' for 800 tweets about the four
presidential candidates Clinton, Cruz, Sanders, and Trump.

We looked for general features that are usually clues for the presence of
sarcasm in a sentence ~\cite{GonzalezMuWa11,DavidovTsRa10} and grouped them into 7 categories:
\begin{enumerate}
\item Quotes: People often copy a candidate's words to make fun of them.
\item Question marks, exclamation or suspension points.
\item All capital letters: Tweeters sometimes highlight sarcasm by writing words
  or whole sentences with all-capital letters.
\item Emoticons like ':)', ':('
\item Words expressing a laugh, or other texting lingo, such as 'ahah,' 'lol,'
  'rofl,' 'OMG,' 'eww,' etc.
\item The words 'yet' and 'sudden.' 
\item Comparisons: Many tweeters use comparisons to make fun of a candidate,
  using words such as 'like' and 'would'.
\end{enumerate}
The sarcasm detecting algorithm that we designed scans the tweet text for those
features and returns the list of sarcastic clues.  The clues are represented by
a 7-component feature vector $f$ that contains a Boolean value for each of the
categories listed above -- ``1'' indicates ``presence'' of the feature, ``0''
otherwise.

Given a tweet $t$ and its feature vector $f$, our method computes the
probability that the tweet~$t$ contains sarcasm by using Bayes rule:
\begin{eqnarray}
P(t \: {\rm is \: sarcastic}|f_n) = \:\:\:\: \:\:\:\:\:\:\:\:\:\:\:\: \:\:\:\:\:\:\:\: \:\:\:\:\:\:\:\: \:\:\:\:\:\:\:\: \:\:\:\:\\
\frac{P(f_n | \, t {\rm \: is \: sarcastic})\: P(t \: {\rm \: is \:
    sarcastic})}{P(f_n)} = \\
 \frac{\# {\rm \: of \: sarcastic \: tweets \: with} \: f_n}{\# {\rm \: of \:
    tweets \: with \: feature} \: f_n}.\:\:\:\: \:\:\:\:\:\:\:\:
\end{eqnarray}
To weigh the presence of the $n$-th feature in sarcastic tweets appropriately,
our method computes a weight vector $w$ by normalizing its $n$-th component by
the probability that it is sarcastic, given any of the seven features is
present:
\begin{equation}
w_n = \frac{P(t \: {\rm is \: sarcastic}|f_n)}{\sum_{n=1}^7 P(t \: {\rm 
is \: sarcastic}|\, f_n)}.
\end{equation}
Our sarcasm score for each tweet is then defined to be the dot product 
\begin{equation} 
w^T \!\! f 
\label{eq:sarcasm-score}
\end{equation} 
of the weight and feature vectors.

\subsection{Decision Tree}
\label{sec:decision-tree}

The decision tree we designed maps a tweet to a number of crowd workers that
will be asked to label the tweet.  To gain insight into the properties of a
tweet that could cause a crowd worker to struggle in sentiment classification
and warrant additional crowd work, we obtained gold standard data and conducted
a formative crowdsourcing study.

\subsubsection{Expert Labels}
We used 1,000 tweets about the four presidential candidates Clinton, Cruz,
Sanders, and Trump.  For these tweets, we had gold standard labels about two
categories, provided by experts in political communication.  The first category
was whether each of the four candidates was mentioned in the tweet.  The second
category described whether the tweet was in general ``positive,'' ``neutral'' or
``negative'' about each candidate mentioned in the tweet.  If more than one
candidate was mentioned in a tweet, the sentiment towards each candidate was
labeled.

\subsubsection{Formative Crowdsouring Experiment} 
%We recruited crowdsourced workers through AMT and accepted only workers based in
%the United States.  We paid workers \$0.02 upon completion of the task.  
We asked 5 crowd workers to analyze each tweet, calling our experiment the
{\em ``Trad 5 baseline''} (the details on the crowdsourcing methodology are given
in Section~\ref{exp-methodology}).  We asked the workers who among the four
candidates Sanders, Trump, Clinton and Cruz was mentioned and to indicate the
attitude that the tweeter expressed towards them on a three-point scale
``positive,'' ``neutral,'' or ``negative.''

%\noindent {\bf 
\subsubsection{Decision Tree Design}
We designed our decision tree (see Fig.~\ref{figure:decision-tree-diagram})
based on the properties we observed that influence the accuracy with which a
worker interprets the sentiment of the tweet.  The first branching of the tree
accounts for whether one or more candidates are mentioned in the tweet text, the
most relevant factor in its sentiment analysis.  Tweets in which several
candidates are mentioned are more difficult to classify because annotators can
become confused by the different attitudes that the writer expresses towards
each of the candidates or by the presence of comparisons between them.  We here
provide three examples:

\noindent
Tweet 1
\begin{quote}
{\em @BecketAdams @JPTruss @GayPatriot except Cruz now realises Trump's power and is
debating him. Rubio is still hiding from Trump on stage}
\end{quote}
is ``positive'' towards Trump and ``neutral'' towards Cruz, according to expert opinion.
Four crowd workers agreed that the message was ``neutral'' towards both candidates,
and one labeled it ``positive'' towards Trump and ``neutral'' towards Cruz.

\noindent Tweet 2
\begin{quote}
{\em Bernie's Super PAC Hypocrisy: Twice as Much Outside Money Spent Supporting
Sanders as Promoting Clinton https://t.co/RVAi7X4shS}
\end{quote}
is ``positive'' towards Clinton and ``negative'' towards Sanders, according to
expert opinion.  All five crowd workers agreed but not on the correct labels --
they selected a negative sentiment towards Sanders and a neutral for Clinton.

\noindent Tweet 3
\begin{quote}
{\em Has Trump mentioned that he doesn't think Cruz is eligible to be President
  recently? That seemed like a go-to for him}
\end{quote}
misled annotators because both sarcasm is present and two candidates are
mentioned. As a consequence, only 3 workers out of 5 agreed on a negative overall
feeling towards both candidates.

It is important whether Clinton or Trump was mentioned in the Tweet.  Opinions
towards these candidates are usually more challenging to understand as tweeters
have very disparate and unclear attitudes towards them.

The next layer of the decision tree accounts for the length of the tweet and the
presence of a link.  We consider a tweet short if it contains fewer than 10
proper words.  Tweets that contain a webpage address are not always fully
understandable by themselves as they refer to the content of the link or they
are a response to another tweet, and therefore their context is not always
clear.

Finally, the terminating decision layer in the tree is based on the sarcastic
score that was produced by the sarcasm predictor.  The decision tree uses the
sarcasm score as defined in Eq.~\ref{eq:sarcasm-score} to determine the
likelihood of sarcasm in the particular tweet.

We assigned a fixed number of crowd workers to each leaf of the tree, which
specifies the number of annotations needed for a particular tweet. In this first
model we grouped the tweets into 4 categories (very easy, easy, medium and hard)
and assigned 2, 3, 5, or 7 workers to them respectively.  We call the model
``Static Decision Tree'' (SDT) due the fact that the number of crowd workers
depends only on the content analysis of the tweet (and not dynamically on the
workers' labels, as described below).  With this tree, the number of crowd
workers to be queried for each tweet can be computed {\em offline} -- in advance
of any crowdsourcing experiment (i.e., the numbers shown in
Fig.~\ref{figure:decision-tree-diagram} with a green-shaded background).

\subsection{Dynamic Worker Assignment}
\label{sec:dynamic-allocation}

We here propose an {\em online} scheme for determining the number of crowd
workers to be queried for each tweet.  This approach cannot be computed in
advance to the crowdsourcing experiment but is an iterative method that relies
on the results of the crowd work.

Our idea is to request a low number of workers to provide the sentiment analysis
of each tweet in a first round of crowdsourcing, and then perform one or more
rounds of crowdsourcing for the tweets for which workers disagreed.  In this
way, the difficulty of the tweet is observed directly as a measure of
disagreement in the first round of crowdsourcing, and we do not risk wasting
effort on tweets that are trivial to classify.  To evaluate our approach, we
designed two instantiations of our idea involving two rounds of crowdsourcing:

\subsubsection{Dynamic Decision Tree 1 (DDT1)}
The first dynamic tree assigns
2 workers to the 'very easy' and 'easy' difficulty classes, 3 for 'medium' and 5
for 'hard.'  If the 2 workers disagree on classifying a 'very easy' or 'easy'
tweet, we conduct a second round of crowdsourcing on that tweet so that we can
get a majority vote.  If some annotators disagree for a 'medium'-class tweet, 2
more workers are involved.  The number of workers for 'hard' tweets stays fixed.

\subsubsection{Dynamic Decision Tree 1 (DDT2)} Finally, we pushed the dynamic
assignment design even further and set up a tree that starts with a very low
numbers of annotators in order to minimize the number of crowdsourced tasks. This
tree initially assigns 2 workers to the 'very easy' and 'easy' classes and
requires 3 more annotators if the initial workers disagree.  The tweets in the
'medium' and 'hard' categories were first only analyzed by 3 workers, and this number
is increased by 2 workers if at least one disagreement is observed.

\subsection{Equivalency of Traditional Static versus Proposed Dynamic Worker Allocation}
\label{sec:off-vs-on}

Past work showed that the probability $p$ that a crowd worker $w$ correctly
performs a task $t$ according to a gold standard label can be described as a
function $p(t,w)$ of the task difficulty and the worker skill ~\cite{HoVa12}.
For simplicity of our analysis, we omit the dependence on the worker.

For a generic task, we can compute the probability~$P_M$ that the gold standard
is successfully obtained by majority voting for a set of crowd sourcing baseline
schemes as a function of $p$. For example, the probability $P_M$ that the
traditional 3-worker-per-task crowdsourcing scheme yields the correct results is
the probability that at least 2 out of 3 performed the task correctly,
which is
\begin{eqnarray}
P_M = \sum_{i=2}^3 P(i  {\rm \, workers \: are \: correct}) = \:\:\:\:\:\:\:\:\:\:\:\:\:\:\:\:\:\:\:\: \nonumber\\
 \sum_{i=2}^3 \binom{3}{i}p^i(1-p)^{(3-i)} = p^2[3(1-p) + p].
\end{eqnarray}
Similarly, with the traditional 5-worker-per-task crowdsourcing scheme, we
attain $P_M=$
\begin{eqnarray}
\label{eq:prob-5}
\sum_{i=3}^5 \! P(i \, {\rm workers \: are \: correct})
\!=
%\:\:\:\:\:\:\:\:\:\:\:\:\:\:\:\:\:\:\:\:  
%\nonumber\\
\! \sum_{i=3}^5 \! \binom{5}{i}p^i(1\!\!-\!\!p)^{(5-i)}   \nonumber \\
= \: p^3[10(1-p)^2+5p(1-p)+p^2]. 
\end{eqnarray}

Next we simulate the dynamic assignment of workers with 3 initial workers,
where 2 more workers are involved if disagreement is encountered.  The
probability that this model produces the correct result by majority voting is
the sum of three probabilities: (1) the probability that the three initial
workers agree on the correct result, (2) the probability that one initial worker
performs the task incorrectly and at least one new worker correctly, and (3) the
probability that only one initial worker performs the task correctly and both
the new workers follow up correctly:
\begin{eqnarray}
\label{eq:prob-dyn}
\binom{3}{3}p^3+\left[\binom{3}{2}p^2(1-p)\right](1-(1-p)^2) \nonumber \\
+\left[\binom{3}{1}p(1-p)^2\right]p^2 = & \nonumber\\
p^3[1+3(1-p)(2-p) + 3(1-p)^2)] =& \nonumber\\
p^3[10(1-p)^2+5p(1-p)+p^2].
\end{eqnarray}

The derivations in Eqs.~\ref{eq:prob-5} and~\ref{eq:prob-dyn} result in the same
formula.  We can therefore infer that a dynamic 3(+2) allocation method for
workers achieves the same prediction accuracy as the traditional 5-worker
crowdsourcing scheme.  As we will describe in more details below, by running
such a model on all tweets in our dataset we were able to obtain optimal results
from crowdsourcing with only 4,058 tasks. This result is impressive because it
proves that we can reach exactly the same accuracy level and save 18.84\% of our
budget only by running two ``smart'' rounds of crowdsourcing.

\section{Experimental Methodology}
\label{exp-methodology}

Our data consists of 1,000 tweets about the four presidential candidates
Clinton, Cruz, Sanders, and Trump sent during the primary election season in
February 2016.  We selected these candidates because they were the two leading
candidates in the polls at the time of data collection from each major
U.S. political party (Republican and Democrat).  The data were collected by
using the Crimson Hexagon ForSight social media analytics platform
(http://www.crimsonhexagon.com/platform).

The tweets were labeled by two domain experts with a background
in political communication in a two-phase process.  In the first phase, the experts
determined the sentiment towards each candidate mentioned in each tweet
independently.  In the second phase, they came to a consensus on the tweets that
they had initially disagreed on.

For our crowdsourcing experiments, we used the Amazon Mechanical Turk (AMT)
Internet marketplace to recruit workers. We accepted all workers from the U.S.
who had previously completed 100 HITs and maintained at least a 92\% approval
rating.  We paid each worker \$0.05 per completed task.  We conducted two
crowdsourcing studies, a formative and a summative study, involving 200 and 800
tweets respectively.

{\bf Formative Study.} We gave the following instruction before presenting
every tweet:
\begin{quote} {\em Carefully read through each tweet and decide the author's attitude
  toward each mentioned presidential candidate (support, neutral, or against).}
\end{quote}
We verified that short tweets (fewer than 10 proper words) were very difficult
to tag.  Tweets with links to an external page were also difficult to analyze.
It is likely that the sentiment of the tweet heavily relies on the content of
the referenced webpage.  Workers may have tried to follow the link or may
have selected a random sentiment instead of following the link.  In our
instructions for our summative study, we therefore specifically asked the crowd
workers not to click on any external link for completing the task.  We also
adjusted the label for positive and negative sentiments towards a candidate.

{\bf Summative Study.} We updated the instructions as follows:
 \begin{quote} {\em Read through the tweet and answer the following questions. Do
    NOT click on any links.    

     Read the tweet and decide whether the candidate was mentioned at all or
     not. Note that the reference of Twitter  user names (e.g.,
     @realDonaldTrump, @HillaryClinton) or hashtags (e.g.,\#Trump2016,
     \#HillaryClinton2016) is also counted as a mention. 
  
     Express which sentiment was manifested by the writer towards them:
     positive, neutral, or negative.}
\end{quote}

We collected ratings from a traditional crowdsourcing scheme that involves 5
independent workers per tweet.  We call this the ``Trad 5'' baseline.  For 15
tweets that were deemed 'hard' to analyze by our decision tree and thus required
the ratings from 7 workers, we needed to collect additional ratings.  Instead of
simply collecting two more, we asked for 5 additional ratings per tweet from
which we could then draw additional samples randomly for analysis.  This resulted
in a total of 5,075 tasks.

To simulate a crowd sourcing experiment that employs a fixed number of three
crowd workers per tweet (our traditional Trad~3 baseline), we randomly sample
the results produced by 5 crowd workers.  To simulate the crowd sourcing
experiments that use the decision tree we designed (SDT, DDT1, DDT2), we
similarly use random samples from our Trad~5 baseline.  To obtain the results of
our decision trees, we averaged the collected metrics over 5 different model
runs to attenuate potential noise generated by the randomness in selecting crowd
workers.

\subsubsection{Evaluation Measures}

We use two metrics for evaluating our work.  They are meaningful for
understanding the trade-off between accuracy and budget concerns, which is the
focus of our work.

\begin{itemize} 
\item {\bf Number of crowd worker tasks:} This is the total number of Human
  Intelligence Tasks requested by our decision tree model. The number provides
  an indication of the budget needs of a crowd experiment. To find the monetary
  costs of crowdsourcing, we can multiply this number by the price per task (we
  used \$0.05/task).

\item {\bf Accuracy of the labeling:} The accuracy of the crowdsourced sentiment
  analysis can be determined by how much agreement exists between the majority
  crowdsourced opinion and the gold standard opinion provided by experts.  Our
  main measure of accuracy is Cohen's Kappa score $\kappa$ for measuring
  inter-rater reliability (IRR).  Cohen's Kappa score accounts for the
  possibility that raters are guessing and so an agreement is obtained by chance.
\end{itemize}

%In order to simulate the performance of the models that we outlined in our
%method we first ran one round of crowdsourcing on the whole dataset of 1000
%tweets on the Amazon Mechanical Turk platform and gathered 5 annotations for
%every tweeter message.  Later we collected 5 more annotations for 15 tweets
%which were labeled as 'hard' by our fixed decision tree and required 7 workers.

\section{Results}

\subsubsection{Sarcasm detection}
Our experiments showed that the clues we used for sarcasm detection are very
diverse, and were used in different ways according to the topic of the tweet.
We found that smileys were not used at all, while the most meaningful element
for sarcasm detection was the presence of expressions like 'lol', 'hahaha,' for
example, in the following tweet:

\begin{quote} {\em RT @rickygervais If Trump was a teacher he'd be fired for
    publicly saying the things he says. Luckily he isn't a teacher. Just the
    next president. Hahaha.}
\end{quote}

The presence of sarcasm was indeed a factor which increased the difficulty of
tweet classification: in our dataset, sarcastic tweets had a 71.2\% percentage
inter-rater agreement.  This metric increased to 78.3\% when dealing with
non-sarcastic tweets.

It turned out that the presence of sarcasm was not as ubiquitous as we had
expected, as only 73 messages out of 800 were estimated to be sarcastic by
domain expert, and a surprising 68.5\% of them concern Donald Trump (see
Table~\ref{table:sarcasm}).  The last row of the table shows that even after
weighing the sarcasm presence over the number of tweets that mentioned each
candidate, Donald Trump still leads with 12\% of his tweets that are sarcastic.
Regarding the sentiment that is usually associated with sarcasm, the last column
of the table proves that sarcasm is usually associated with a negative feeling
towards a candidate. In fact this language feature is usually employed to make
fun of a candidate and criticize him for his statements or actions.

\begin{table}
\small
\centering
\begin{tabular}{l||c|c|c|c|r}
& \bf Clinton & \bf Cruz & \bf Sanders & \bf Trump & \\
\hline \hline
Positive & 0 & 0 & 2 & 3 & 5 \\
Neutral & 5 & 6 & 4 & 15 & 30 \\
Negative & 6 & 4 & 6 & 32 & 48 \\
\hline
& 11 & 10 & 12 & 50 & \\
Sarcastic & 5.9\% & 6.2\% & 6.8\% & 12.0 \% &
\end{tabular}
\caption{
These results show the number of sarcastic tweets
  addressed to each candidate and  the  sentiment that they  showed according to the gold
  standard provided by experts in political communication.
  In the dataset of 800 tweets, 73 tweets were sarcastic.
  The last row shows the ratio of
  sarcastic tweets over the total tweets in which each candidate was mentioned.}
\label{table:sarcasm} 
\end{table}

\subsubsection{Differences Based on Specific Candidates}
As expected, we found that which presidential candidate was mentioned in a tweet
had an impact on how difficult it was to discern the tweeter's opinion about the
candidate.  The sentiments that tweeters expressed towards Hillary Clinton and
Donald Trump were often unclear or veiled by sarcasm.  To illustrate this point
qualitatively, we give an example tweet about Trump that confused the crowd
workers:
\begin{quote} {\em I was watching the Texas gop debate on snapchat lol and this
    is the only state where I've seen people actually rally against trump YOUNG
    PPL.}
\end{quote}
One crowd worker labeled the tweet to show ``a positive attitude,'' 2 crowd
workers labeled it as ``neutral'' and the remaining 2 agreed on a ``negative''
sentiment towards the candidate.  In this case, it is impossible to determine a
result by majority vote, and a final label can be assigned by a reasonable
random choice.  We here chose randomly between ``neutral'' and ``negative.''

To illustrate the issue quantitatively, we here provide the inter-rater
reliability values among 5 crowd workers of our formative study when classifying
sentiments towards each candidate and report both the relative observed
agreement among crowd workers and Cohen's Kappa score $\kappa$: \vspace{0.2cm}

\begin{tabular}{lcr}
 Candidate & Agreement & Kappa IRR\\
 Bernie Sanders: & 83.05\%  & $\kappa$ = 0.74 \\
 Ted Cruz:  & 87.78\%   & $\kappa$ = 0.78 \\
 Hillary Clinton: & 63.41\%  & $\kappa$ = 0.41  \\
 Donald Trump  & 78.13\% & $\kappa$ = 0.66\\
\end{tabular}

\vspace{0.2cm}

It is evident from the above numbers that annotators disagreed much more often
when Clinton or Trump were mentioned.  For our summative study, we therefore
designed an offline model that can account for this observation and involve more
workers to label tweets from these two candidates.

% MEHRNOOSH:
%1. As we would like to reduce disagreement by involving more people, the reviewers might ask why we hypothesized that the disagreements might come from the difficulty of our dataset and not inefficiency of our task design/guidelines? Why didn't we consider changing our instructions or task design to improve disagreements? I suggest adding a section where we explain the process we took to come up with our final interface and task design. we first conducted experiments with the first 200 tweets and learned a lot of details about our task improvement (we got the IRR's for before and after those improvements and stopped modifying the task as soon as IRR among workers and experts were higher than 0.7). For instance, we added this sentence to the final instructions base on our experience in round 1: "Do not click on any links." We also limited our workers to US workers.

\begin{table}
\small
\centering
\begin{tabular}{l||l|l||r|r|r}
& \bf Trad 3 & \bf Trad 5 & \bf SDT & \bf DDT1 & \bf DDT2  \\ 
\hline
Efficiency & 3,000 & 5,000 & 3,907 & 3,206 & 3,608\\
\hspace*{0.2cm} Imprv. & & & 22\% & 36\% & 28\% \\ 
Accuracy   & 0.612 & 0.653 & 0.624 & 0.630 & 0.643\\
\hspace*{0.2cm} Loss & & & 4.4 pp & 3.5 pp & 1.0 pp\\ 
\hline
\end{tabular}
\caption{
  Comparison of results of five methods with
  respect to their efficiency and accuracy.    The number of crowd workers
  engaged (i.e., efficiency or costs) and the accuracy of their sentiment
  labeling (Cohen's Kappa IRR rate) compared to the gold standard established by experts are given
  for each method.   For the first two methods, each
  tweet is analyzed by the same fixed number of crowd workers, i.e.,  3
  crowd workers (Trad 3) or 5 crowd workers (Trad 5).  For the methods that use the
  decision tree (DT), the number of crowd workers engaged depends on the
  content of the tweet and result in significant improvements (Improv.) in
  efficiency with respect to the 5
  crowd-worker models (row 2), without much loss of accuracy (row 4, given in
  percent points, pp).
\label{table:results} 
}
\end{table}

\subsubsection{Results for Traditional Fixed-Allocation Model} 

The first two models that we considered are a fixed crowdsourcing round with the
same amount of workers for every tweet. With a total of 3 annotators we
requested 3,000 ratings and we achieved a 0.612 Kappa value (see
Table~\ref{table:results}).  If we increase the number of crowd workers by 2 we
require 5,000 tasks and we would get a 0.653 reliability measure. These results
align with previous observations that the task of sentiment analysis is
challenging even for human annotators~\cite{YoungSo12,TumasjanSpSaWe10} Despite
the significantly higher costs of requesting 2,000 additional labels from crowd
workers, a 40\% increase, the average agreement between the majority of crowd
contributions and expert labels improved by only 6.3 percent (or, equivalently,
by a difference of Kappa values of 4.1 percent points).

\subsubsection{Results for the Proposed Static Decision Tree} 

For the static decision tree (SDT), 3,907 labels were requested, on average, and
an IRR score of 0.624 was obtained.  The allocated numbers of workers based on
the text analysis of the tweets and decision rules of the tree are shown in red
in Figure~\ref{figure:decision-tree-diagram}.  With this static decision tree,
22\% of the budget would be saved with respect to the traditional
5-worker-per-task model (Trad 5).  The loss in accuracy is 4.4 percent points.

\subsubsection{Results for the Proposed Dynamic Decision Trees} 

The first dynamic tree (DDT1) showed a meaningful improvement as it involves
only 3,206 tasks on average and has an IRR score of 0.630.  This model costs
36\% less than the fixed one with 5 workers and only 6.9\% more than the model
with 3 annotators but the gain in accuracy with respect to the latter is quite
high (2.9\%). This model would be preferable in low-budget situations.

The second dynamic tree (DDT2) is a bit more expensive as it requires 3,608
annotators by average but the Cohen's Kappa IRR rate improves to 0.643. Even
this classifier is much cheaper than the fixed 5-worker as it saves almost 28\%
of the budget and the accuracy is comparable (the difference between Kappas
scores is only 1 percent point). We propose that this predictor is suitable if
we are willing to spend a bit more in order to achieve a very good performance.

Both dynamic trees produce notably better results than the fixed decision tree
in both cost and accuracy. This shows that the difficulty of a tweet can be
inferred from the crowdsourcing outcomes themselves and that heuristic rules for
determining it are extremely complex and hard to formulate.  Correct results can
be obtained by a second round of annotations, which needs to be set up
accordingly, thus saving a meaningful amount of budget.

\subsubsection{Cost Savings of Dynamic versus Static Worker Assignment}

\begin{figure*}[t]
  \centering
  \includegraphics[width=0.9\linewidth]{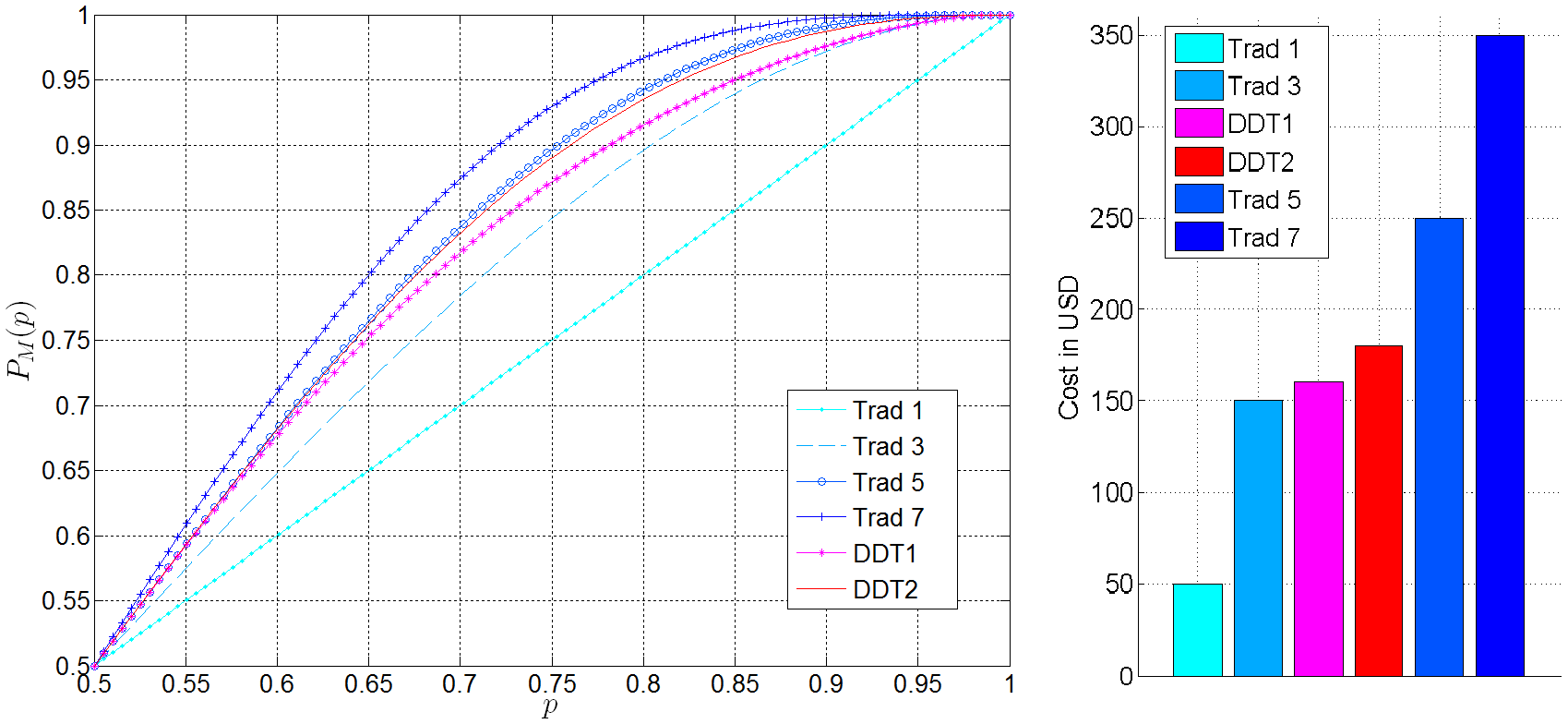}
  \caption{Performance Analysis -- Accuracy and Costs. Left: The probability
    $P_M(p)$ that a given crowdsourcing scheme produces the correct label by
    majority vote as a function of the probability that a certain tweet is
    labeled correctly by a worker.  We compare the performance of four
    traditional crowdsourcing baselines (with 1, 3, 5 or 7 crowd workers for
    each tweet) and our dynamic prediction models DDT1 and DDT2.  For tweets
    that are easy to annotate, the accuracy of all methods is similar.  When
    tweets are more difficult to analyze, and thus, more workers are engaged, the
    performance gains in accuracy of the DDT1 and DDT2 models compared to the
    traditional models ``Trad~3'' become apparent.  The DDT2 model almost
    reaches the performance of the baseline ``Trad~5.''  Right: The proposed
    dynamic models DDT1 and DDT2 provide large budget savings.}
\label{figure:prob}
\end{figure*}

The traditional 5-worker-per-task allocation model Trad 5 performs exactly the
same as a dynamic model which assigns 3 annotators +2 more if there is
disagreement, as described in Section~\ref{sec:off-vs-on}.  This result shows
that our model allows the same accuracy but at a much lower cost.  A
visualization of the differences in accuracy and efficiency between traditional
static crowdsourcing schemes and the proposed dynamic schemes is given in
Figure~\ref{figure:prob}.

\subsubsection{Analysis of Crowd Work Properties}

We submitted 5,075 tasks to Mechanical Turk for an overall cost of \$253.75.
The number of MT workers who contributed labels to all the tweets was 218.  An
average of 23 annotations was submitted per worker.

\begin{figure*}[t]
  \centering
  \includegraphics[width=0.9\linewidth]{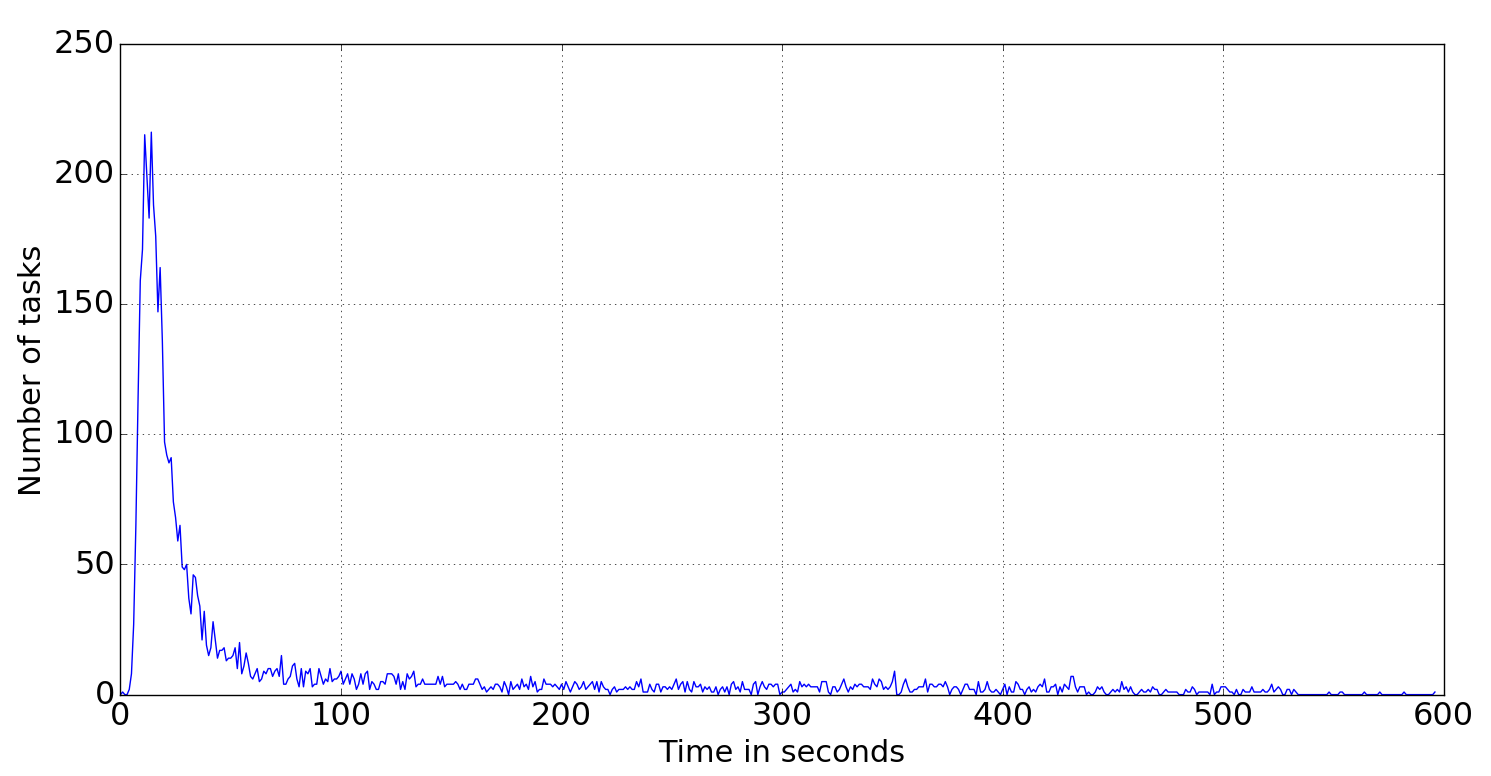}
  \caption{A distribution of tasks (HITs) as a function of task time, ranging
    from 1 to 600 seconds. This distribution was computed over the total 5,075
    tasks that were submitted to Amazon Mechanical Turk during our crowdsourcing
    experiment.}
\label{figure:time}
\end{figure*}

We analyzed how much time workers spent in labeling a single tweet, which is
illustrated in Figure~\ref{figure:time}. Annotators spent an average of 85.1
seconds for classifying a single message but some workers were very meticulous
and used up to 10 minutes to complete a single task.  For example one of the
best annotators who worked for us labeled 217 tweets with an average of 212
seconds per task, which sums up to almost 13 hours spent on the platform. On the
other hand, other annotators were very quick, for instance one worker contributed
by labeling 42 tweets and spent on average less than 9 seconds per message.

\subsubsection{Sample Results on Political Tweets}

Analysis of the annotations of our 1,000 tweet dataset provides some fascinating
observations about political opinions.  We can report the overall sentiment that
people showed towards candidates, as rated by the crowd workers
(Table~\ref{table:crowdsentiment}) and by the experts in political communication
(Table~\ref{table:expertsentiment}).  We found that Trump is the ``most
popular'' candidate to tweet about, considering that more than half of the total
tweets mentioned him, while the other candidates were evenly referred to on
average. Furthermore it is clear that tweeters who discuss candidates for
presidential elections often express negative feelings and complain about
candidates, since there are about twice as many negative messages than positive
ones in our entire dataset.  The main difference between the crowd worker and
expert annotations was the tendency of the crowd worker to label fewer tweets as
``neutral.''

\begin{table}
\small
\centering
\begin{tabular}{l||c|c|c|c|r}
& \bf Clinton & \bf Cruz & \bf Sanders & \bf Trump & \\
\hline \hline
Positive & 33 & 44 & 78 & 123 & 278 \\
Neutral & 86 & 67 & 104 & 189 & 446 \\
Negative & 99 & 99 & 56 & 201 & 455 \\
\hline
& 218 & 210 & 238 & 513 &
\end{tabular}
\caption{
  Number of tweets, out of a total of 800, grouped according to crowd-sourced sentiment label per candidate.
  The last row and columns display the sums over the columns and rows respectively
  in the table.}
\label{table:crowdsentiment}
\end{table}

\begin{table}
\small
{\centering
\begin{tabular}{l||c|c|c|c|r}
& \bf Clinton & \bf Cruz & \bf Sanders & \bf Trump  & \\
\hline \hline
Positive  & 25 & 37 & 58 & 90 & 210 \\
Neutral  & 109 & 85 & 123 & 208 & 525 \\
Negative  & 91 & 89 & 55 & 212 & 447 \\
\hline
& 225 & 211 & 236 & 510 &
\end{tabular}}
\caption{
  Number of tweets, out of a total of 800, grouped according to expert-provided
  sentiment label per candidate.
  The last row and columns display the sums over the columns and rows respectively
  in the table.}
\label{table:expertsentiment} 
\end{table}

\section{Discussion and Conclusions}

As crowdsourcing becomes more and more popular for large scale information
retrieval, the cost of this human computation is becoming relevant.  Example
applications are real-time sentiment analysis to provide fast indications of
changes in public opinion or collection of a sufficiently large training data
for machine learning methods for big data analytics~\cite{WangCaKaBaNa12}.
Investigations, as ours, about how to balance the goals of efficiency and
accuracy in crowdsourcing, are therefore particularly timely.

Few works have explored dynamic approaches to crowdsourcing that rely on
iterative rounds of crowdsourcing and determine the number of worker assignments
based on content and annotation results in previous
rounds~\cite{BraggKoMaWe14,HoVa12,KolobovMaWe13}. Connections to active and
reactive learning ~\cite{YanRoFuDy11,LinMaWe15} have been made.  While prior
work involves theoretical analysis and simulation studies, we here provide a
concrete solution to the problem of analyzing the sentiment of political twitter
messages using a dynamic worker allocation framework.

We proposed a dynamic two-round crowdsourcing scheme that we embedded into a
decision tree classifier.  Other types of classifiers may be used, and, in
future work, we will explore additional learning methods.

Analysis of political tweets is challenging due to the short text and unknown
context.  Sentiment analysis is particularly difficult. Existing off-the-shelf
text analysis systems can only provide a single sentiment label for a given text
automatically.  We found that they fail to distinguish the separate sentiments
that were expressed when more than one presidential candidate was mentioned in a
tweet.  The presence of sarcasm exacerbated the problem.  Our proposed solution
is to design a classifier that early in the analysis makes a decision about the
number of sentiments that must be revealed.  Our new dataset may inspire other
researchers to develop text analysis tools that address the difficult problem of
multi-sentiment analysis and sarcasm detection.

%We are committed to sharing our dataset and source code at {\tt
%  http://anonymous.edu}.  
Our corpus of 1,000 twitter messages is unique because
it includes information about (1) the presence/absence of sarcasm and (2) a
label about the specific sentiment for each candidate mentioned in the tweet
(positive, neutral, negative), as determined by consensus of two domain experts.

It is notable that our study involved communication researchers in many aspects
of the research, such as the development and refinement of crowdsourcing task
instructions and the design of the Mechanical Turk interface. The intervention
of domain experts greatly helped improve the validity and performance of our
crowdsourcing method.

Likewise, the proposed approach has the potential to make a significant
contribution to communication research. Traditionally, communication researchers
use manual content analysis, a method that usually relies on two or three human
coders, to analyze text in different media outlets or that of public
opinion~\cite{RiffeLaFi14}. However, the traditional method is tedious, time
consuming, and limited by the nature of human subjectivity. Arguably, the use of
the dynamic online crowdsourcing framework introduced in this study allows
communication researchers to process larger datasets in a more efficient and
reliable manner.  Given the results of the study, future research should also
consider cross-disciplinary collaboration to advance theories and methods for
large-scale text analysis.
 
\section*{Acknowledgments}
The authors would like to thank the Boston University Rafik B. Hariri Institute
for Computing and Computational Science and Engineering for financial support
and the crowd workers for their annotations.

\bibliography{crowdsource}
\bibliographystyle{aaai}

\end{document}